\documentstyle[aps,prb,epsf]{revtex}

\begin{document}
\twocolumn[\hsize\textwidth\columnwidth\hsize\csname
@twocolumnfalse\endcsname
\title{Spectral function and quasiparticle weight in the generalized $t-J$ model}
\author{F. Lema and A. A. Aligia}
\address{Centro At\'{o}mico Bariloche and Instituto Balseiro,\\
Comisi\'on Nacional de Energ\'{\i}a At\'omica,\\
8400 Bariloche, Argentina}
\maketitle

\begin{abstract}
We extend to the spectral function an approach which allowed us to calculate
the quasiparticle weight for destruction of a real electron $Z_{c\sigma }(%
{\bf k)}$ (in contrast to that of creation of a spinless holon $Z_{h}({\bf k)%
}$) in a generalized $t-J$ model, using the self-consistent Born
approximation (SCBA). We compare our results with those obtained using the
alternative approach of Sushkov {\it et al.,} which also uses the SCBA. The
results for $Z_{c\sigma }({\bf k)}$ are also compared with results obtained
using the string picture and with exact diagonalizations of a 32-site square
cluster. While on a qualitative level, all results look similar, our SCBA
approach seems to compare better with the ED one. The effect of hopping
beyond nearest neighbors, and that of the three-site term are discussed.
\end{abstract}


\vskip2pc]

\narrowtext

\section{Introduction}

The momentum distribution function of holes in a quantum antiferromagnetic
background has been a subject of considerable interest since the discovery
of high-$T_{c}$ systems. Reliable information on these quantities was
obtained from the exact diagonalization of the $t-J$ model in small clusters 
\cite{szc,s2,s3,poi,leu}. An analytical approximation which brings
considerable insight into the underlying physics is based on the string
picture \cite{ede}. Within the string picture, for realistic $t>J$, the
movement of the hole can be separated into a fast motion around a fixed
position $j$ on the lattice (to which the hole is attracted by a string
potential caused by the distortion of the Neel background), and a slow
motion of $j$ due essentially to spin fluctuations which restore the Neel
background as $j$ is displaced. The resulting quasiparticle weight as a
function of wave vector $Z_{c\sigma }({\bf k)}$ for one hole, agrees very
well with exact diagonalizations (ED) of a $4\times 4$ cluster \cite{ede}.
However, this cluster is still too small and finite-size effects are
important \cite{poi,leu}.

Another successful analytical approach is the self-consistent Born
approximation (SCBA) \cite{sch,kan,mar,liu,bal}. The resulting dispersion of
one hole in the $t-J$ model is in very good agreement with exact
diagonalizations of small clusters \cite{leu,mar,liu}, including a square
cluster of 32 sites, which contains the most important symmetry points and
smaller finite-size effects in comparison with previous calculations \cite
{leu}. The SCBA has also been used to calculate the doping dependence of the
superconducting critical temperature $T_{c}$ in qualitative agreement with
experiment \cite{pla}. However, due to the particular representation used
(see Eq. (\ref{f2}) below), the Green function which results from the SCBA
is that of a spinless holon $G_{h}$, while the real particle becomes related
with a composite operator $c_{k\sigma }$ (composed of a holon and a spin
deviation). The holon quasiparticle weights $Z_{h}({\bf k)}$ differ from $%
Z_{c\sigma }({\bf k)}$ which are the physical quantities calculated by ED
and accessible to experiment. Only recently, motivated by photoemission
experiments in insulating Sr$_{2}$CuO$_{2}$Cl$_{2}$ \cite{wel}, two
approaches appeared which relate $Z_{c\sigma }$ \cite{lem} and the Green
function of the physical hole $G_{c\sigma }$ \cite{sus} with $G_{h}$ within
the framework of the SCBA. In addition, to fit accurately the experimentally
observed dispersion, it is necessary to include second- $(t_{2})$ and
third-nearest-neighbor hopping $(t_{3})$ to the $t-J$ model \cite{bel,xia},
and to explain qualitatively the observed intensities, it is necessary to
consider the strong-coupling limit of a Hubbard model \cite{lem,esk}. This
implies that a three-site term $t^{\prime \prime }=-J/4$ should be included
in the model, and the one-particle operators should be transformed. While a
generalized Hubbard model is able to provide a consistent picture of the
observed charge and and spin excitations in Sr$_{2}$CuO$_{2}$Cl$_{2}$ \cite
{lem2}, we must warn that the effective strong-coupling low-energy effective
model derived from a realistic multiband model \cite{fei,f2,ali,a2},
although also includes $t_{2}$ and $t_{3}$, can have a different $t^{\prime
\prime }$, even of opposite sign, favoring instead of suppressing of $d$%
-wave superconductivity and a resonance-valence-bond ground state \cite
{rvb,r1,r2,r3}.

For future theoretical studies, as well as to compare with photoemission 
\cite{wel,mars,pot}, or other experiments (like electron-energy-loss
spectroscopy \cite{wan,bos}) it is important to compare the four above
mentioned approaches to calculate the quasiparticle dispersion and
intensities, trying to establish their relative accuracy or convenience.
This is the main purpose of this work.

In Section 2 we derive an expression for the spectral density of the
physical hole $\rho _{c\sigma }=-$Im$G_{c\sigma }/\pi $ in terms of the
Green function of the spinless hole $G_{h}$ within the SCBA, and briefly
describe the alternative expression derived by Sushkov {\it et al.} \cite
{sus} for $G_{c\sigma }$. Section 3 contains a comparison of both resulting $%
\rho _{c\sigma }({\bf k,}\omega )$ after solving the SCBA equations. In
Section 4, we compare the results for $Z_{c\sigma }({\bf k)}$ obtained using
both SCBA approaches with ED results for the $t-J$ model \cite{leu}, and
those derived from a string picture \cite{ede}. In Section 5 we include $%
t_{2}$, $t_{3}$ and $t^{\prime \prime }$ and compare recent ED results for
the 32-site cluster \cite{leu2} with the SCBA ones. Section 6 contains the
conclusions.

\section{Green function of a physical hole within the SCBA}

We consider the dynamics of one hole in a square lattice described by a
generalized $t-J$ model: 
\begin{eqnarray}
H &=&-\sum_{i\delta \sigma }t_{\delta }c_{i+\delta \sigma }^{\dag
}c_{i\sigma }+{\frac{J}{2}}\sum_{i\eta }({\bf S}_{i}\cdot {\bf S}_{i+\eta }-{%
\frac{1}{4}}n_{i}n_{i+\eta })  \nonumber \\
&&+t^{\prime \prime }\sum_{i\eta \neq \eta ^{\prime }\sigma }c_{i+\eta
^{\prime }\sigma }^{\dag }c_{i+\eta \sigma }({\frac{1}{2}}-2{\bf S}_{i}\cdot 
{\bf S}_{i+\eta }).  \label{f1}
\end{eqnarray}
The first term contains hopping to first, second and third nearest neighbors
with parameters $t_{1},~t_{2},~t_{3}$ respectively. The nearest neighbors of
site $i$ are labeled as $i+\eta $. In the SCBA, long-range antiferromagnetic
order is assumed, and a spinless fermion $h_{i}$ at each site is introduced 
\cite{sch,kan,mar,liu,bal}. Calling A (B) the sublattice of positive
(negative) spin projections in the Neel state, one can use the following
representation:

\begin{eqnarray}
c_{i\uparrow } &=&h_{i}^{\dagger },\,\,\,\,\,\,\,\,\,c_{i\downarrow
}=h_{i}^{\dagger }a_{i}\,,\,\,\,\,\,\text{if }\,\,i\in \,\text{A}  \nonumber
\\
c_{i\uparrow } &=&h_{i}^{\dagger }a_{i},\,\,\,c_{i\downarrow
}=h_{i}^{\dagger }\,\,,\,\,\,\,\,\,\,\,\,\,\text{if}\,\,\,\,i\in \,\text{B,}
\label{f2}
\end{eqnarray}
where $a_{i}^{\dagger }$ creates a spin deviation at site $i$ in the Neel
state, and there is a constraint that at the same site there cannot be both,
a hole and a spin deviation. This constraint is neglected \cite{mar}. The
exchange part of Eq. (1) is diagonalized by a standard canonical
transformation, and retaining only linear terms in the spin deviations for
the other terms, the Hamiltonian takes the form:

\begin{eqnarray}
H &=&\sum_{q}\omega _{q}\alpha _{q}^{\dagger }\alpha _{q}+\sum_{k}\epsilon
_{k}h_{k}^{\dagger }h_{k}  \nonumber \\
&&+\frac{4t_{1}}{\sqrt{N}}\sum_{kq}M({\bf k,q})(h_{k}^{\dagger
}h_{k-q}\alpha _{q}+\text{H.c.}),  \label{f3}
\end{eqnarray}
where $\alpha _{q}=u_{q}a_{q}-v_{q}a_{q}^{\dagger }$, $\omega _{q}=2J\,\nu
_{q},$ $\epsilon _{k}=(t_{2}+2(1-x)t^{\prime \prime })\epsilon
_{2}(k)+(t_{3}+(1-x)t^{\prime \prime })\epsilon _{1}(2k)$ $x=1/N$ is the
doping, $\epsilon _{1}(k)=4\gamma _{k}$, $\epsilon _{2}(k)=4$ cos $k_{x}$
cos $k_{y}$,

\begin{eqnarray}
v_{q}^{2} &=&u_{q}^{2}-1=\frac{1}{2\nu _{q}}-\frac{1}{2},\ \nu
_{q}=(1-\gamma _{q}^{2})^{1/2},  \nonumber \\
\gamma _{q} &=&(\cos q_{x}+\cos q_{y})/2,\ \text{sgn}(v_{q})=\text{sgn}%
(\gamma _{q}),  \nonumber \\
M({\bf k,q}) &=&u_{q}\gamma _{k-q}+v_{q}\gamma _{k},  \label{f4}
\end{eqnarray}
and $u_{q}>0$.

From the Hamiltonian Eq.(\ref{f3}), the SCBA allows to calculate the holon
Green function $G_{h}$ accurately through the self-consistent solution of
the following two equations:

\begin{eqnarray}
\Sigma ({\bf k},\omega ) &=&\frac{z^{2}t_{1}^{2}}{N}\sum_{q}M^{2}({\bf k,q}%
)G_{h}({\bf k-q},\omega -\omega _{q})  \nonumber \\
\,\,\,\,\,\,\,\,\,\,\,\,\,\,\,\,\,\,\,\,G_{h}^{-1}({\bf k},\omega )
&=&\omega -\epsilon _{k}-\Sigma ({\bf k},\omega )+i\epsilon .  \label{f5}
\end{eqnarray}
$z=4$ is the coordination number.

However, the physical operator is the Fourier transform of $c_{i\sigma }$
and since it is a composite operator in the representation Eq.(\ref{f2}), it
is not trivial to find its Green function $G_{c\sigma }$. Recently, two
different approaches to calculate the quasiparticle weight $Z_{c\sigma }(%
{\bf k)}$ \cite{lem} and the whole Green function $G_{c\sigma }({\bf k}%
,\omega )$ \cite{sus} using the SCBA have been proposed. Here we extend to
the spectral density $\rho _{c\sigma }({\bf k,}\omega )=-$Im$G_{c\sigma }(%
{\bf k,}\omega )/\pi $ our previous derivation \cite{lem}. We assume for the
moment that the system is finite and its eigenvalues are discrete. The idea
is to relate $G_{c\sigma }$ and $G_{h}$, using the equation of motion \cite
{rei} to find the wave function for each eigenvector. If $|\psi
_{k}^{n}\rangle $ is an eigenstate of Eq. (\ref{f3}) with total wave vector 
{\bf k }and other quantum numbers labeled by $n$, it can be expanded as:

\begin{eqnarray}
|\psi _{k}^{n}\rangle &=&A_{0}^{n}({\bf k})h_{k}^{\dagger }|0\rangle +\frac{1%
}{\sqrt{N}}\sum_{q}A_{1}^{n}({\bf k,q})h_{k-q}^{\dagger }\alpha
_{q}^{\dagger }|0\rangle  \nonumber \\
&&+\frac{1}{N}\sum_{q_{1},q_{2}}A_{2}^{n}({\bf k,q}_{1},{\bf q}%
_{2})h_{k-q_{1}-q_{2}}^{\dagger }\alpha _{q_{2}}^{\dagger }\alpha
_{q_{1}}^{\dagger }|0\rangle +...,  \label{f6}
\end{eqnarray}
and the Schr\"{o}dinger equation $(\lambda _{k}^{n}-H)|\psi _{k}^{n}\rangle
=0$, leads to an infinite set of equations for the $A_{i}^{n}$. Neglecting
some terms not described by the SCBA \cite{rei}, the first two of these
equations are:

\begin{eqnarray}
A_{0}^{n}({\bf k})(\lambda _{k}^{n}-\epsilon _{k}) =\frac{1}{N}%
\sum_{q}A_{1}^{n}({\bf k,q})M({\bf k,q})  \nonumber \\
A_{1}^{n}({\bf k,q})(\lambda _{k}^{n}-\epsilon _{k-q}-\omega _{q}) \simeq
zt_{1}[A_{0}^{n}({\bf k})M({\bf k,q})  \nonumber \\
+\frac{1}{N}\sum_{q_{2}}A_{2}^{n}({\bf k,q},{\bf q}_{2})M({\bf k-q,q}%
_{2})].  \label{f7}
\end{eqnarray}
A solution of the set of approximate equations can be solved relating $%
A_{i}^{n}$ with $A_{i-1}^{n}$ \cite{rei}. In particular, if

\begin{eqnarray}
A_{2}^{n}({\bf k,q},{\bf q}_{2}) &=&zt_{1}A_{1}^{n}({\bf k,q})M({\bf k-q,q}%
_{2})  \nonumber \\
&&\times G_{h}({\bf k-q-q}_{2},\lambda _{k}^{n}-\omega _{q}-\omega _{q_{2}}),
\label{f8}
\end{eqnarray}
with $G_{h}({\bf k},\omega )$ and $\Sigma ({\bf k},\omega )$ satisfying Eqs.
(\ref{f5}), from Eqs. (\ref{f7}) it follows that:

\begin{equation}
A_{1}^{n}({\bf k,q})=zt_{1}A_{0}^{n}({\bf k})M({\bf k,q})G_{h}({\bf k-q}%
,\lambda _{k}^{n}-\omega _{q}),  \label{f9}
\end{equation}
and the eigenvalue equation:

\begin{equation}
A_{0}^{n}({\bf k})[\lambda _{k}^{n}-\epsilon _{k}-\Sigma ({\bf k},\lambda
_{k}^{n})]=0\text{.}  \label{f10}
\end{equation}
Since we are considering the case of only one hole at zero temperature
(otherwise Eqs. (\ref{f5}) have to be generalized \cite{mah,tes}), the
relevant Fock space is composed by $|0\rangle $ (the ground state of the
Heisenberg antiferromagnet), and the eigenstates with one added hole $|\psi
_{k}^{n}\rangle $. In this restricted Hilbert space, the Lehmann
representations \cite{mah,tes} of the relevant Green functions read:

\begin{eqnarray}
G_{c\sigma }({\bf k},\omega ) &=&\sum_{n}\frac{|\langle \psi
_{k}^{n}|c_{k\sigma }|0\rangle |^{2}+|\langle \psi _{k+(\pi ,\pi
)}^{n}|c_{k\sigma }|0\rangle |^{2}}{\omega -\lambda _{k}^{n}+i\epsilon }, 
\nonumber \\
G_{h}({\bf k},\omega ) &=&\sum_{n}\frac{|\langle \psi
_{k}^{n}|h_{k}^{\dagger }|0\rangle |^{2}}{\omega -\lambda _{k}^{n}+i\epsilon 
}.  \label{f11}
\end{eqnarray}
Using Eqs. (\ref{f2}) and the expression of the $a_{i}$ in terms of the
magnon operators $\alpha _{q}$ to Fourier transform $c_{i\sigma }$ \cite{lem}%
, Eqs.(\ref{f6}), (\ref{f9}), (\ref{f11}), and some algebra, we find for the
ratio of spectral functions:

\begin{equation}
\frac{\rho _{c\sigma }({\bf k,}\omega )}{\rho _{h}({\bf k,}\omega )}=\frac{1%
}{2}|1+\frac{2zt_{1}}{N}\sum_{q}^{\prime }v_{q}M({\bf k,q})G_{h}({\bf k-q}%
,\omega -\omega _{q})|^{2},  \label{f12}
\end{equation}
where the sum runs over all wave vectors of the antiferromagnetic Brillouin
zone, excluding ${\bf q=0}$, and $v_{q}$ and $M({\bf k,q})$ are given by Eq.
(\ref{f4}).

Although Eq. (\ref{f12}) has been derived assuming that the eigenvalues are
discrete, we expect it to be also valid for a continuum distribution of
energy levels (the incoherent background of the spectrum), or in other
words, when the small imaginary part $i\epsilon $ in Eqs. (\ref{f5}) and (%
\ref{f12}) is larger than the average spacing between the levels. This seems
to be confirmed by the results shown in the next section. A similar method
of using the Schr\"{o}dinger equation assuming discrete eigenvalues was used
to obtain the non-interacting Green function of the Anderson model \cite{ham}%
.

Sushkov {\it et al.} \cite{sus} used a different method to relate $%
G_{c\sigma }({\bf k},\omega )$ with $G_{h}({\bf k},\omega )$. They treat the
operators $c_{k\sigma }$ as originated by an external perturbation to the
Hamiltonian Eq. (\ref{f3}). This perturbation is characterized by two
vertices in which the physical hole is transformed into a spinless hole $%
h_{k}^{\dagger }$, with or without exchange of a magnon. The physical Green
function $G_{c\sigma }({\bf k},\omega )$ is obtained from a Dyson equation.
Another difference with our approach lies in the different normalization of
the operators $h_{k}^{\dagger }$, which is equivalent to another way of
treating the constraint that there cannot be a hole and a spin deviation at
the same site. We have neglected it since it has been shown that it does not
affect the results for a quantum antiferromagnet (the situation is different
if the spins are described by an Ising model) \cite{mar}. Sushkov {\it et
al. }also have a factor $\sqrt{2}$ in their normalization of the $c_{k\sigma
}$, which introduces a factor 2 in $G_{c\sigma }$. Dropping this factor to
compare with other results, their final expression for the physical Green
function reads:

\begin{eqnarray}
G_{c\sigma }({\bf k},\omega ) =0.4G_{h}({\bf k},\omega )+\frac{1}{N}%
\sum_{q}^{\prime }v_{q}^{2}G_{h}({\bf k-q},\omega -\omega _{q})  \nonumber \\
+4\sqrt{0.8}t_{1}G_{h}({\bf k},\omega )\frac{1}{N}\sum_{q}^{\prime }v_{q}M(%
{\bf k,q})G_{h}({\bf k-q},\omega -\omega _{q})  \nonumber \\
+32t_{1}^{2}G_{h}({\bf k},\omega )[\frac{1}{N}\sum_{q}^{\prime }v_{q}M(%
{\bf k,q})G_{h}({\bf k-q},\omega -\omega _{q})]^{2},  \label{f13}
\end{eqnarray}
where $v_{q}$ and $M({\bf k,q})$ are given by Eq. (\ref{f4}).

\section{Comparison of spectral densities within SCBA.}

The expression of the spectral density of the physical hole $\rho _{c\sigma
}({\bf k,}\omega )$ derived from Eq. (\ref{f13}) is quite different from Eq.
(\ref{f12}). Nevertheless, in both cases it is necessary first to solve
numerically the SCBA Eqs. (\ref{f5}) to obtain the Green function of the
spinless hole $G_{h}({\bf k},\omega )$. To perform this task with high
accuracy, we have discretized the frequencies in intervals of $\Delta \omega
=10^{-4}t_{1}$ and have taken the small imaginary part as $\epsilon =5\Delta
\omega $. We have chosen a cluster of $16\times 16$ sites.

\begin{figure}
\narrowtext
\epsfxsize=3.0truein
\vbox{\hskip 0.05truein \epsffile{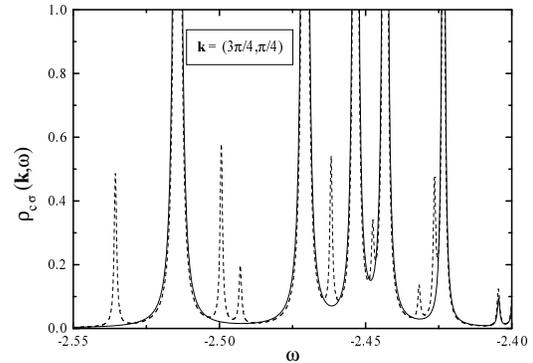}}
\medskip
\caption{Spectral function of the physical hole obtained from
the SCBA for wave vector {\bf k}$=(3\pi /4,\pi /4)$. Full line: result of
Eq.(12). Dashed line: result derived from the Dyson Eq. (13) [13]. 
Parameters are $t_{1}=1,~J=0.3,~t_{2}=-0.3,$ $t_{3}=0.2,$ 
$t^{\prime \prime }=0$.}
\end{figure}

\medskip

\begin{figure}
\narrowtext
\epsfxsize=3.0truein
\vbox{\hskip 0.05truein \epsffile{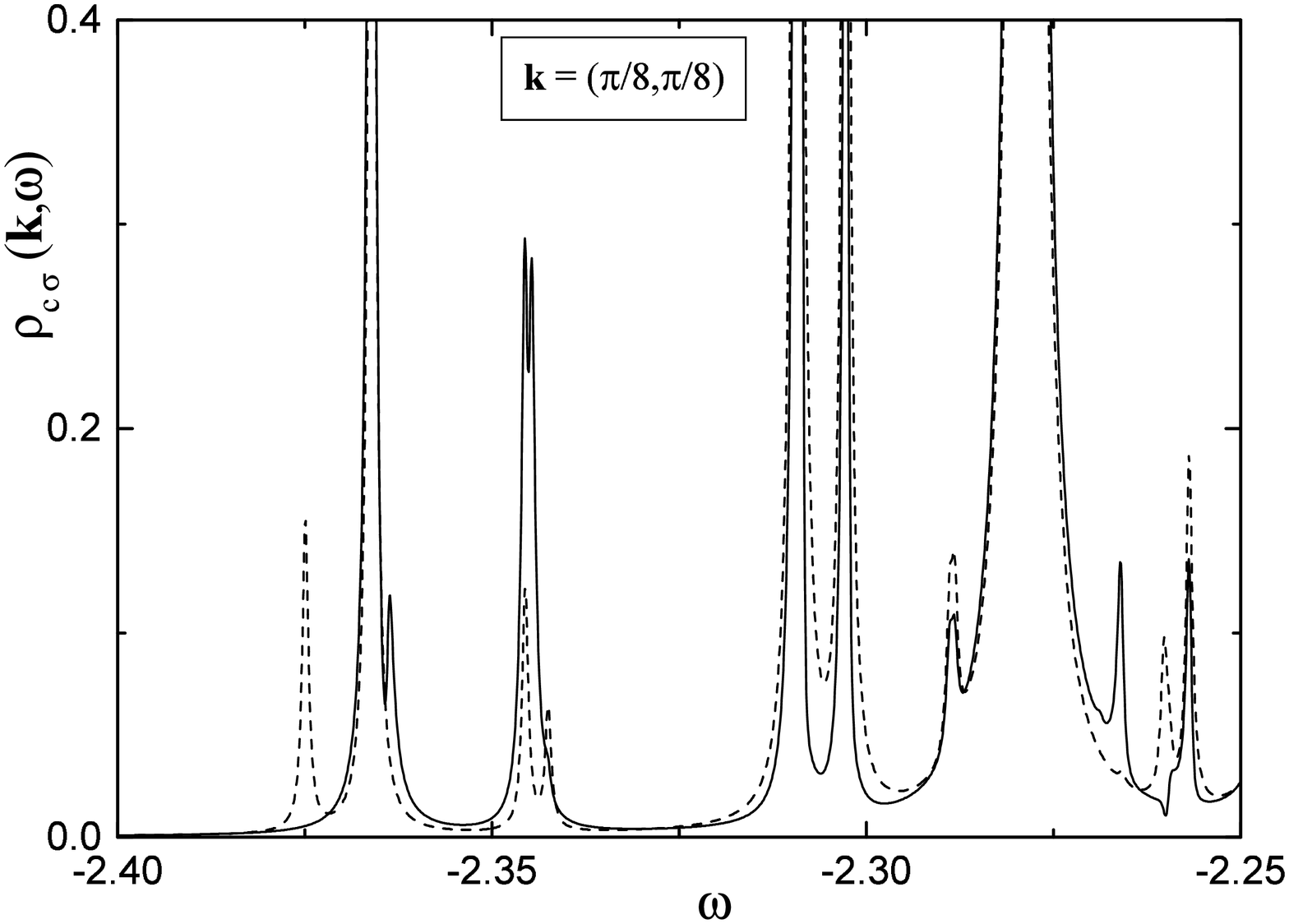}}
\medskip
\caption{Same as Fig. 1 for {\bf k}$=(\pi /8,\pi /8)$.}
\end{figure}

In Figs. 1 and 2, we represent the resulting spectral densities for a wave
vector {\bf k} at the boundary of the antiferromagnetic Brillouin zone, and
another one near the zone center respectively. Except for wave vectors near $%
(\pi ,\pi )$ at frequencies near the quasiparticle pole, the contribution of
the term in $t_{1}$ in Eq. (\ref{f12}) is small compared to 1, and as a
rough first approximation Eq. (\ref{f12}) gives $\rho _{c\sigma }({\bf k,}%
\omega )\simeq \rho _{h}({\bf k,}\omega )/2$. Thus, the result of our Eq. (%
\ref{f12}) looks qualitatively similar to the known results for $\rho _{h}$
within the SCBA \cite{mar,liu}. There are several peaks which can be
qualitatively understood within the string picture as originated by
different bound states of the string potential, which acquire dispersion as
the center of the string potential is displaced by spin fluctuations or
terms of sixth order in $t_{1}$ \cite{ede,mar}. The lowest peak corresponds
to the coherent quasiparticle state. Note that the structure in the
incoherent background displaying several different peaks cannot be resolved
if the imaginary part in Eqs. (\ref{f5}) is as large as that used by Sushkov 
{\it et al.} ($\epsilon =0.1t_{1}$ \cite{sus}). The same happens for certain
wave vectors and parameters with the quasiparticle peaks, when the
quasiparticle energy $\lambda _{k}$ lies too near the incoherent background.

The result for $\rho _{c\sigma }({\bf k,}\omega )$ obtained using the Dyson
Eq. (\ref{f13}) derived by Sushkov {\it et al.} \cite{sus} looks in general,
and for any reasonable parameters of the generalized $t-J$ model, similar to
ours, except for two differences: i) the intensity is a little bit smaller
for the peaks already present in the spinless holon result $\rho _{h}({\bf k,%
}\omega )$ (see Fif. 1), ii) new peaks appear, which apparently do not have
a physical meaning and seem to be an artifact of the approximations involved
in the Dyson equation. In particular, except for high symmetry points of the
Brillouin zone (like $(0,0)$, $(0,\pi )$ or $(\pi ,\pi )$), for wave vectors 
{\bf k} such that the quasiparticle energy $\lambda _{k}$ is near $\lambda
_{(0,0)}$ (the bottom of the electron band), a new quasiparticle peak
appears. Comparison with the position and intensity of the quasiparticle
peak at {\bf k} $=(3\pi /4,\pi /4)$ obtained from exact diagonalization of a
square 32-site cluster \cite{leu,leu2}, described in the following two
sections, confirm that this peak is spurious and is disregarded in the
following.

\section{Quasiparticle weights in the $t-J$ model}

To obtain the quasiparticle intensity for each wave vector of the physical
hole $Z_{c\sigma }({\bf k})$ within the SCBA, we have fitted the part of the
spectral density (described by Eq. (\ref{f12}) or Eq. (\ref{f13}) and Eq. (%
\ref{f4})) in the neighborhood of the quasiparticle peak by a sum of several
Lorentzian functions. Since the imaginary part $\epsilon =5\times
10^{-4}t_{1}$ we have taken in the numerical solution of the SCBA Eqs. (\ref
{f5}) is very small, we can isolate the quasiparticle peak (its width is
practically identical to $2\epsilon $), even in some cases where the
distance of this peak to the incoherent background is smaller than $\epsilon 
$. Integrating the corresponding Lorentzian we obtain $Z_{c\sigma }({\bf k})$%
. We have verified that using this method, finite-size effects are
practically absent in our $16\times 16$ cluster. The peaks introduced by Eq.
(\ref{f13}) which are absent in the spinless hole spectral density $\rho
_{h}({\bf k,}\omega )$ were neglected.

\begin{figure}
\narrowtext
\epsfxsize=3.3truein
\vbox{\hskip 0.05truein \epsffile{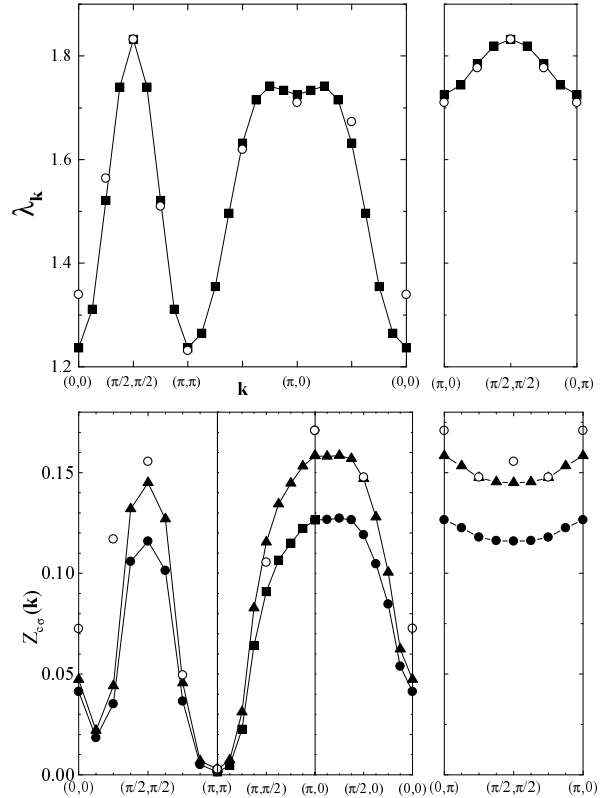}}
\medskip
\caption{Top: quasiparticle energies obtained with the SCBA
(solid squares). Bottom: corresponding quasiparticle weights obtained using
Eq.(12) [12] (solid triangles) and from Dyson Eq. (13) [13] (solid circles).
Open circles denote the corresponding results of
an exact diagonalization (ED) of a square cluster of 32 sites [3].
Parameters are $t_{1}=1,~J=0.3,~t_{2}=t_{3}=t^{\prime \prime }=0$.}
\end{figure}

In Fig. 3 we compare quasiparticle energies $\lambda _{k}$ and weights $%
Z_{c\sigma }({\bf k})$ with those obtained from an exact diagonalization
(ED) of a square cluster of 32 sites \cite{leu}. The SCBA results for $%
\lambda _{k}$ were already published \cite{liu}. We inverted the sign of $%
\lambda _{k}$ (the electron instead of the hole representation is used) in
the following, in order to facilitate comparison with previous calculations
and experiment. We also shifted the $\lambda _{k}$ in order that they
coincide for the ground-state wave vector ${\bf k}=(\pi /2,\pi /2)$, as in
Ref. \cite{leu}. The agreement between ED and SCBA results for $\lambda _{k} 
$ is excellent along the boundary of the antiferromagnetic Brillouin zone.
However, there are some discrepancies near the zone center, particularly for 
${\bf k}=(0,0)$ and ${\bf k}=(\pi /4,\pi /4)$. We ascribe this to
finite-size effects, since in the thermodynamic limit $\lambda _{k}=\lambda
_{k+(\pi ,\pi )}$, due to the folding of the Brillouin zone caused by the
antiferromagnetic symmetry breaking. Except for the above mentioned two wave
vectors, the disagreement between the results for $Z_{c\sigma }({\bf k})$ of
the ED and our SCBA approach \cite{lem} (Eq. (\ref{f12})) is less than 7\%.
Using instead the SCBA expression (\ref{f13}) of Sushkov {\it et al.} \cite
{sus}, we obtain a quasiparticle weight which is $\sim 20\%$ below our
results. Previous comparison of ED results for $Z_{c\sigma }({\bf k})$ on a
square cluster of 20 sites \cite{poi} and SCBA results on a $20\times 20$
cluster using Eq. (\ref{f12}), also agreed very well except at ${\bf k}%
=(0,0) $.

\begin{figure}
\narrowtext
\epsfxsize=3.3truein
\vbox{\hskip 0.05truein \epsffile{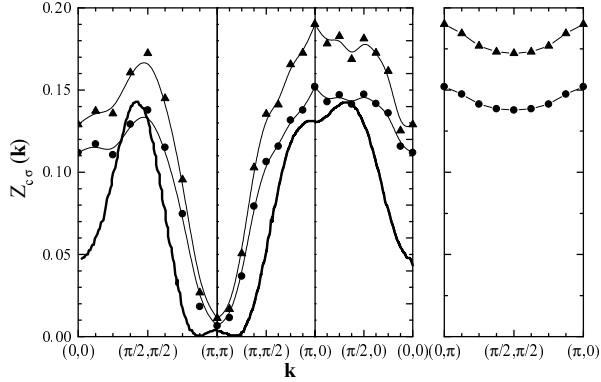}}
\medskip
\caption{Quasiparticle weights obtained with the SCBA using
Eq.(12) [12] (solid triangles) and Eq. (13) [13] (solid circles). 
The full thick line corresponds to results of the string
picture scanned from Ref. 4. Parameters are 
$t_{1}=1,~J=0.4,~t_{2}=t_{3}=t^{\prime \prime }=0$.}
\end{figure}

Comparison between results of ED for clusters between 16 and 32 sites \cite
{poi,leu}, suggest that finite-size effects are very important for the $%
4\times 4$ cluster. In Fig. 4 we compare the results for $Z_{c\sigma }({\bf k%
})$ obtained using the string picture, scanned from Ref. \cite{ede} with
both SCBA approaches. ED results on large enough clusters are not available
for these parameters. While the three curves look qualitatively similar, it
seems that the string picture results underestimate $Z_{c\sigma }({\bf k})$,
particularly at ${\bf k}=(0,0)$ and near ${\bf k}=(\pi ,\pi )$.

\section{Quasiparticle energies and weights in the generalized $t-J$ model}

In this Section, we compare results for $\lambda _{k}$ and $Z_{c\sigma }(%
{\bf k})$ which we obtained using the SCBA, with corresponding recent ED
results, in which hoppings beyond nearest neighbors $t_{2},$ $t_{3}$ and the
three-site term $t^{\prime \prime }$ were included \cite{leu2}. A motivation
to include these terms is that the inclusion of $t_{2}$ and $t_{3}$ is
necessary \cite{bel,xia} to explain the experimentally observed dispersion
in insulating Sr$_{2}$CuO$_{2}$Cl$_{2}$ \cite{wel}. However, to explain
qualitatively the observed quasiparticle intensities, it is necessary to
include at least Hubbard corrections to the relevant operators \cite
{lem,sus,esk}. Their effect can be included in any theoretical approach and
its main effect is to increase the intensities near the Brillouin zone
center \cite{lem,sus,esk}.

\begin{figure}
\narrowtext
\epsfxsize=3.3truein
\vbox{\hskip 0.05truein \epsffile{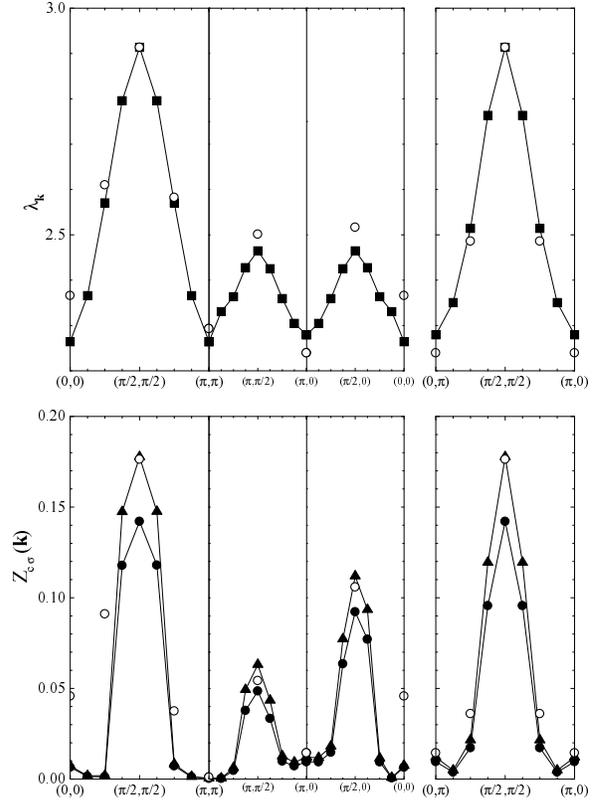}}
\medskip
\caption{Same as Fig. 3 for the parameters of Fig. 1. The ED
results were taken from Ref. 25.}
\end{figure}

Fig. 5 contains the results for $t^{\prime \prime }=0$. Comparing with Fig.
3 one can see that the effect of $t_{2}$ and $t_{3}$ is mainly to shift $%
\lambda _{(\pi ,0)}$ towards the incoherent background, and as a
consequence, the weight $Z_{c\sigma }(\pi ,0)$ is strongly reduced. The same
happens for neighboring {\bf k}. Also, the agreement between ED and SCBA
results for $\lambda _{k}$ near ${\bf k}=(\pi ,0)$ is not so good as for $%
t_{2}=t_{3}=0$. In spite of this, the agreement between ED and SCBA results
for $Z_{c\sigma }({\bf k})$ is still very good except at the points ${\bf k}%
=(0,0)$, ${\bf k}=(\pi /4,\pi /4)$ and ${\bf k}=(3\pi /4,3\pi /4)$, for
which finite-size effects are present in the ED calculations as evidenced by
the fact that $\lambda _{k}\neq \lambda _{k+(\pi ,\pi )}$.
In Fig. 6, we include the three-site term $t^{\prime \prime }=-J/4$, with
magnitude corresponding to the strong-coupling limit of the Hubbard model.
Its effect is to lower the quasiparticle energies $\lambda _{k}$ near the
wave vectors $(0,0)$ and $(\pi ,0)$, increasing the total dispersion. While
the comparison between SCBA and ED results for $\lambda _{k}$ is not
affected appreciably by $t^{\prime \prime }$, the corresponding weights $%
Z_{c\sigma }({\bf k})$ clearly disagree for wave vectors $(\pi /2,0)$ and ($%
\pi ,\pi /2)$. However, the agreement between weights on the boundaries of
the antiferromagnetic Brillouin zone continues to be satisfactory.

\begin{figure}
\narrowtext
\epsfxsize=3.3truein
\vbox{\hskip 0.05truein \epsffile{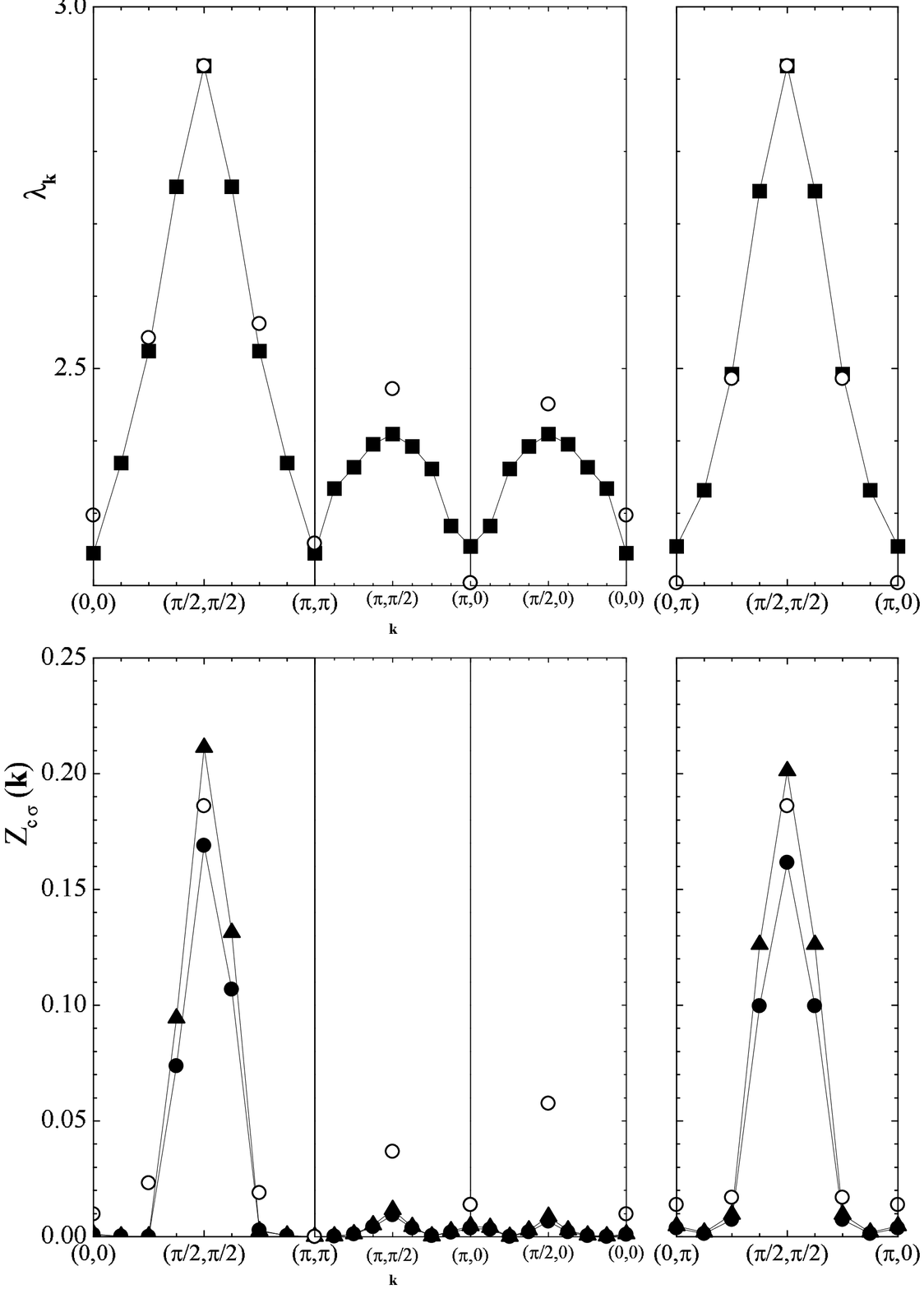}}
\medskip
\caption{Same as Fig. 5 for $t^{\prime \prime }=-J/4$.}
\end{figure}

\section{Summary and discussion}

We have derived an expression which relates the spectral density of a
physical hole with spin $\rho _{c\sigma }({\bf k,}\omega )$, with that of a
spinless hole $\rho _{h}({\bf k,}\omega )$, and calculated it within the
self-consistent Born approximation (SCBA). Another approach for the Green
function of the physical hole, proposed by Sushkov {\it et al.} \cite{sus},
introduces additional structure and spurious quasiparticle peaks in $\rho
_{c\sigma }({\bf k,}\omega )$ for some wave vectors {\bf k}. However, they
can be identified and eliminated by comparison with the results obtained for 
$\rho _{h}$ within the SCBA.

We have also compared the SCBA quasiparticle energies $\lambda _{k}$ and
weights $Z_{c\sigma }({\bf k})$ with available exact-diagonalization (ED)
results on a square cluster of 32 sites for different parameters \cite
{leu,leu2}, and with calculations using the string picture \cite{ede}. On a
qualitative level, both SCBA approaches \cite{lem,sus}, the string picture,
and the ED results are similar. Quantitatively, although the string picture
is very useful to give insight into the underlying physics, it seems that it
underestimates the weights. The SCBA approach of Sushkov {\it et al.} \cite
{sus} looks more accurate, but for all parameters and wave vectors studied
here, the resulting $Z_{c\sigma }({\bf k})$ is smaller than the
corresponding ED result and our SCBA one \cite{lem}.

The agreement between $Z_{c\sigma }({\bf k})$ obtained from ED and our SCBA
approach is very good, particularly for the $t-J$ model except at two {\bf k 
}points where finite-size effects are evident in the ED results, and on the
boundary of the antiferromagnetic Brillouin zone. In general, the $%
Z_{c\sigma }({\bf k})$ are larger if the corresponding $\lambda _{k}$ are
near $\lambda _{(\pi /2,\pi /2)}$ (which correspond to the ground state for
all parameters studied here). Inclusion of other hopping terms shifts $%
\lambda _{k}$, particularly for ${\bf k}\sim (\pi ,0)$ away from $\lambda
_{(\pi /2,\pi /2)}$, the corresponding weights $Z_{c\sigma }({\bf k})$
decrease, and the agreement between SCBA and ED results becomes poor for
some of these wave vectors. This disagreement might be due to the fact that
a nearest-neighbor hopping of the chosen sign weakens the antiferromagnetic
long-range order \cite{ali,a2,af,af2,af3}. While the existence of this
order is an essential assumption of the SCBA, and is true for one hole in an
infinite system, it is destroyed for a small finite concentration of holes $%
\sim 0.015 $ in cuprates \cite{exp}. This concentration is smaller than the
corresponding one (1/32) in the ED studies. Another possibility for the
discrepancies is that vertex corrections which seem to not affect
essentially the spinless hole Green function $G_{h}$ \cite{bal}, become
important for $Z_{c\sigma }({\bf k})$, particularly when the three-site term 
$t^{\prime \prime }$ is present.

For a quantitative comparison with photoemission experiments, it is
necessary to transform the relevant operators of the appropriate multiband
model, involving oxygen and copper orbitals, into those of the effective
generalized $t-J$ model. This task might be performed generalizing previous
related studies \cite{bat,fei,f2,sim}.

\section*{Acknowledgments}

One of us (FL) is supported by the Consejo Nacional de Investigaciones
Cient\'{\i }ficas y T\'{e}cnicas (CONICET), Argentina. (AAA) is partially
supported by CONICET.

\end{document}